\documentclass[dvips,12pt,a4paper]{article}
\usepackage{a4wide}
\usepackage{epsfig}
\usepackage{amsmath}
\usepackage{latexsym}
\usepackage{array}

  \newlength{\absize}
\newcommand{\dd}{\mbox{{\rm d}}}

\newcommand{\Lumint}{{\cal L}_{\rm int}}

\allowdisplaybreaks 

\catcode`@=11
\def\citer{\@ifnextchar [{\@tempswatrue\@citexr}{\@tempswafalse\@citexr[]}}

\def\@citexr[#1]#2{\if@filesw\immediate\write\@auxout{\string\citation{#2}}\fi
  \def\@citea{}\@cite{\@for\@citeb:=#2\do
    {\@citea\def\@citea{--\penalty\@m}\@ifundefined
       {b@\@citeb}{{\bf ?}\@warning
       {Citation `\@citeb' on page \thepage \space undefined}}%
\hbox{\csname b@\@citeb\endcsname}}}{#1}}
\catcode`@=12

\begin{document}
  \thispagestyle{empty}
  \pagestyle{empty}
  \renewcommand{\thefootnote}{\fnsymbol{footnote}}
\newpage\normalsize
    \pagestyle{plain}
    \setlength{\baselineskip}{4ex}\par
    \setcounter{footnote}{0}
    \renewcommand{\thefootnote}{\arabic{footnote}}
\newcommand{\preprint}[1]{%
  \begin{flushright}
    \setlength{\baselineskip}{3ex} #1
  \end{flushright}}
\renewcommand{\title}[1]{%
  \begin{center}
    \LARGE #1
  \end{center}\par}
\renewcommand{\author}[1]{%
  \vspace{2ex}
  {\Large
   \begin{center}
     \setlength{\baselineskip}{3ex} #1 \par
   \end{center}}}
\renewcommand{\thanks}[1]{\footnote{#1}}
\renewcommand{\abstract}[1]{%
  \vspace{2ex}
  \normalsize
  \begin{center}
    \centerline{\bf Abstract}\par
    \vspace{2ex}
    \parbox{\absize}{#1\setlength{\baselineskip}{2.5ex}\par}
  \end{center}}

\begin{flushright}
{\setlength{\baselineskip}{2ex}\par

}
\end{flushright}
\vspace*{4mm}
\vfill
\title{Model-independent constraints on contact interactions \\ 
from LEP2 data analysis}
\vfill
\author{
A.A. Babich$^{a,b}$, G. Della Ricca$^{c}$, J. Holt$^{d}$, \\ 
P. Osland$^{e}$, A.A. Pankov$^{a,c}$ {\rm and}
N. Paver$^{c,}$\footnote{E-mail address: nello.paver@ts.infn.it}}
\begin{center}
$^a$ Pavel Sukhoi Technical University, 
     Gomel 246746, Belarus \\
$^b$ ICTP, 34014 Trieste, Italy \\
$^c$ University of Trieste  and INFN-Sezione di Trieste, 34100 
Trieste, Italy \\
$^d$ CERN, 1211 Geneva 23, Switzerland\\
$^e$ Department of Physics, University of Bergen, 
N-5007 Bergen, Norway 
\end{center}
\vfill 
\abstract {We derive model-independent constraints on four-fermion
contact interaction-type dynamics from the published preliminary LEP2
experimental data on $e^+e^-$ annihilation into $\mu^+\mu^-$ and
$\tau^+\tau^-$ pairs, measured at different energies between 130 and
207~GeV. The basic observables are chosen to be the total cross section and
the forward-backward asymmetry, and the analysis realistically takes into
account data uncertainties and correlations among measurements at the various
energies.  The combination of data from different energy points plays an
important role in the determination of regions allowed for the contact
interaction coupling constants.
In contrast to the more common one-parameter analyses, we only obtain
constraints on pairs of parameters rather than limits on individual ones.}

\vspace*{20mm}
\setcounter{footnote}{0}
\vfill

\newpage
    \setcounter{footnote}{0}
    \renewcommand{\thefootnote}{\arabic{footnote}}
\section{Introduction}
Many Standard Model extensions envisage a dynamics acting at one (or more)
large mass scales $\Lambda\gg M_{W}$, such that the relevant states exchanged
among quarks and leptons, having a mass proportional to $\Lambda$, are so
heavy that they could not be directly produced at accelerator energies. The
most familiar case is represented by quark and lepton composite models
\cite{'tHooft:xb,Eichten:1983hw}, but there are numerous other examples.
However, such new interactions could manifest themselves by indirect, virtual,
effects represented by deviations of the measured observables from the
Standard Model (SM) predictions.  If some deviations were effectively observed
experimentally to a given significance level, one could try to derive from the
data numerical information on the parameters (masses and coupling constants)
of the non-standard models and, eventually, to select the viable one. In the
case where, instead, no deviation from the SM predictions were observed within
the experimental accuracy, one can set numerical bounds and/or constraints on
the parameters characterizing the new interactions and, in particular, on the
relevant mass scales $\Lambda$. This information should also be of
phenomenological interest, in the exploration of non-standard interactions.
\par
In the spirit of ``effective'' theories, exchanges of very heavy objects in
reactions of quarks and leptons can be parameterized by a {\it contact
interaction}, representing the ``low energy'' expansion of the transition
amplitude to leading order in the small ratio $\sqrt s/\Lambda$ ($\sqrt s$
being the c.m.\ energy). The explicit form of such contact interaction
Lagrangian (CI) depends on the particles participating in the reaction under
consideration. Specifically, we consider here the electron-positron
annihilation:
\begin{equation}
e^++e^-\to f+\bar{f}, \label{proc} \end{equation}   
with $f=\mu$ and $\tau$, and the relevant precision data at LEP2 for 
$130< \sqrt s <207$ GeV, published in Ref.~\cite{geweniger}, where the 
results of the four experimental collaborations are combined. Such 
high precision data can be regarded as a powerful tool to severely test 
manifestations of non-standard interactions through 
deviations from the SM predictions. In particular, we are interested in 
deriving, from those data, constraints on the {\em e e f f}
{\it contact-interaction} Lagrangian \cite{Eichten:1983hw}:
\begin{equation}
{\cal L}=\sum_{\alpha\beta}{g^2_{\rm eff}}\hskip2pt
\epsilon_{\alpha\beta}\left(\bar e_{\alpha}\gamma_\mu e_{\alpha}\right)
\left(\bar f_{\beta}\gamma^\mu f_{\beta}\right),
\label{lagra}
\end{equation}
where $\alpha,\beta={\rm L,R}$ denote left- or right-handed fermion 
helicities, and the parameters 
$\epsilon_{\alpha\beta}$ specify the chiral structure of individual
interactions and determine the size of the deviations from the SM
predictions. One can introduce the previously mentioned large mass scales
by $\vert\epsilon_{\alpha\beta}\vert=1/\Lambda_{\alpha\beta}^2$,
and coventionally fixing $g_{\rm eff}^2/{4\pi}=1$ as a reminder that,
as a compositeness remnant force, this interaction would become strong
at $\sqrt{s}\sim\Lambda_{\alpha\beta}$. However, as 
remarked above, more generally the scales $\Lambda_{\alpha\beta}$ 
define a standard to compare the sensitivity of measurements to the 
various kinds of new interactions, see, {\it e.g.}, 
\cite{Barger:1997nf,Altarelli:1997ce}. 
\par
In practice, the situation is complicated by the fact that, for a given 
fermion flavor $f$, Eq.~(\ref{lagra}) defines four individual and 
independent models (basically, the combinations of the four chiralities
$\alpha,\beta$ through the $\epsilon$'s) and, in principle,
the general contact interaction could be any linear combination of these 
models. Thus, the aforementioned deviations of the cross 
section from the SM predictions may simultaneously depend on {\it all} 
four-fermion effective couplings and, if only one value of the c.m.\ energy  
were available, the straightforward comparison of deviations and 
experimental uncertainty could produce only numerical correlations among 
the the different CI couplings, rather than separate, and restricted, 
allowed regions for these parameters in the parameter space around the 
SM limit $\epsilon_{\alpha\beta}=0$. Moreover, negative interference 
of CI and SM amplitudes in the cross section might considerably weaken 
the bounds. 
\par
The simplest and commonly adopted procedure consists in assuming non-zero
values for just one of the $\epsilon_{\alpha\beta}$ at a time, and in
constraining it to a finite interval by essentially a $\chi^2$ fit analysis of
the measured cross sections and forward-backward asymmetries, while all the
other parameters are set equal to zero \citer{Barger:1997nf,Kroha:1991mn}. In
this way, only tests of specific models can be performed.
\par 
On the other hand, it would be desirable to 
perform a more general kind of analysis of the data, that
simultaneously includes all terms of Eq.~(\ref{lagra}) as free, potentially 
non-vanishing independent parameters and, at the same time, allows to 
disentangle their contributions to the basic observables in order to 
derive separate constraints within finite regions around the SM limit. 
\par
In cases where only one value for the c.m. energy is  
available, such as for the planned $e^+e^-$ Linear Collider 
\cite{Aguilar-Saavedra:2001rg}, a solution is represented by the initial 
electron beam's longitudinal polarization, that would enable to
experimentally extract the individual helicity amplitudes of process 
(\ref{proc}), by definition directly related to the individual 
{\em e e f f} contact couplings $\epsilon_{\alpha\beta}$ 
\cite{Babich:1999kp,Babich:2001nc}. 
\par 
Such a procedure cannot be applied to the data from LEP, with unpolarized 
electron and positron beams. However, in this case, the cross sections 
of processes (\ref{proc}) are measured at LEP2 over a range of $\sqrt s$ 
values wide enough that the energy dependence of the deviations, entirely  
determined by well-known SM parameters, can be exploited to restrict the 
bounds to limited regions in the CI parameter space, and in this way to 
perform an analysis of the new interaction, model-independent in the 
sense indicated above. This observation was used for a global 
analysis of data at the energies of LEP1, LEP2 and TRISTAN in 
Ref.~\cite{Pankov:2000jf}. The analysis presented here uses exclusively the 
most recent higher statistics LEP2 data, combines the two 
channels $\mu^+\mu^-$ and $\tau^+\tau^-$ and the results of the four 
experiments, and accounts for, among other things, the correlations 
among the measurements at the different energy points.
The basic observables will be the ``conventional'' ones, namely, the 
integrated cross section $\sigma (s)$ and the forward-backward
asymmetry $A_{\rm FB}(s)$, whose experimental values are tabulated 
in Ref.~\cite{geweniger}. 
\par 
Specifically, in Sec.~2 we will give the basic definition of helicity 
amplitudes and the formulae relevant to $\sigma$ and $A_{\rm FB}$ for 
the processes of interest here, and in Sec.~3 we shall present the 
model-independent analysis of LEP2 data and the resulting constraints 
on CI couplings. Finally, Sec.~4 will be devoted to some concluding 
remarks and an application of the method to a model example. 

\section{Cross section and helicity amplitudes}
Limiting ourselves to the cases $f=\mu, \tau$ and neglecting all fermion 
masses with respect to $\sqrt s$, and taking into account the 
Born $\gamma$ and $Z$ exchanges in the $s$ channel plus the
{\it contact-interaction} term (\ref{lagra}), the differential cross 
section of process (\ref{proc}) reads \cite{Schrempp:1987zy}: 
\begin{equation}
\frac{\dd\sigma}{\dd\cos\theta}
=\frac{3}{8}
\left[(1+\cos\theta)^2 {\sigma}_+
+(1-\cos\theta)^2 {\sigma}_-\right], 
\label{cross}
\end{equation}
where $\theta$ is the angle between the incoming electron and the 
outgoing fermion in the c.m. frame. In terms of helicity cross sections,  
$\sigma_{\alpha\beta}$ with $\alpha,\beta={\rm L,R}$:
\begin{eqnarray}   \label{Eq:sigma+}
{\sigma}_{+}&=&\frac{1}{4}\,
\left(\sigma_{\rm LL}+\sigma_{\rm RR}\right),
\label{s+} \\
   \label{Eq:sigma-}
{\sigma}_{-}&=&\frac{1}{4}\,
\left(\sigma_{\rm LR}+\sigma_{\rm RL}\right).
\label{s-}
\end{eqnarray}
In Eqs.~(\ref{s+}) and (\ref{s-}):
\begin{equation}
\sigma_{\alpha\beta}=\sigma_{\rm pt}
\vert {\cal M}_{\alpha\beta}\vert^2,
\label{helcross}
\end{equation}
where $\sigma_{\rm pt}\equiv\sigma(e^+e^-\to\gamma^\ast\to l^+l^-)
=4\pi\alpha_{\rm e.m.}^2/3s$ (for quark-antiquark production a color 
factor $N_C\simeq 3(1+\alpha_s/\pi)$ would be needed). The helicity 
amplitudes ${\cal M}_{\alpha\beta}$ can be written as
\begin{equation}
{\cal M}_{\alpha\beta}=Q_e Q_f+g_\alpha^e\,g_\beta^f\,\chi_Z+
\frac{s}{\alpha_{\rm e.m.}}\epsilon_{\alpha\beta},
\label{amplit}
\end{equation}
where: $\chi_Z=s/(s-M^2_Z+iM_Z\Gamma_Z)$ is the $Z$ propagator; $g_{\rm
L}^f=(I_{3L}^f-Q_f s_W^2)/s_W c_W$ and $g_{\rm R}^f=-Q_f s_W^2/s_W c_W$ are
the SM left- and right-handed fermion couplings of the $Z$ with
$s_W^2=1-c_W^2\equiv \sin^2\theta_W$; $Q_e=Q_f=-1$ are the fermion electric
charges.  
\par 
The measured observables $\sigma$ and $A_{\rm FB}$ are given by the relations:
\begin{equation}
\label{sigma}
\sigma=\int_{-1}^1 \frac{\dd\sigma}{\dd\cos\theta}\,\dd\cos\theta
=\frac{1}{4}\left[\left(\sigma_{\rm LL}+\sigma_{\rm RR}\right)
+\left(\sigma_{\rm LR}+\sigma_{\rm RL}\right)\right];
\end{equation}
and
\begin{equation}
\sigma_{\rm FB}\equiv\sigma\, A_{\rm FB}
=\left(\int_{0}^1-\int_{-1}^0\right) \frac{\dd\sigma}{\dd\cos\theta}\,
\dd\cos\theta
=\frac{3}{16}\left[\left(\sigma_{\rm LL}+\sigma_{\rm RR}\right)
-\left(\sigma_{\rm LR}+\sigma_{\rm RL}\right)\right].
\label{sigmafb}
\end{equation}
Finally, their relation to $\sigma_{\pm}$ is given by 
\begin{equation}
\sigma_{\pm}=
\frac{\sigma}{2}\hskip 2pt \left(1\pm\frac{4}{3}\hskip 1pt A_{\rm FB}\right).
\label{sig+-}
\end{equation}
\par 
Taking Eq.~(\ref{amplit}) into account, Eqs.~(\ref{sigma}) and 
(\ref{sigmafb}) show that $\sigma$ and $\sigma_{\rm FB}$ (or $A_{\rm FB}$)  
simultaneously depend on {\it all} four contact interaction couplings, and 
therefore by themselves do not allow a
model-independent analysis, but only the simplified one-parameter fit of
individual models. However, $\sigma$ and $\sigma_{\rm FB}$ depend on the 
two combinations of helicity cross sections, 
$\left(\sigma_{\rm LL}+\sigma_{\rm RR}\right)$ and 
$\left(\sigma_{\rm LR}+\sigma_{\rm RL}\right)$. Accordingly, a combined 
analysis of $\sigma$ and $\sigma_{\rm FB}$ enables to separately constrain 
the pairs of parameters $(\epsilon_{\rm LL},\epsilon_{\rm RR})$ and 
$(\epsilon_{\rm LR},\epsilon_{\rm RL})$. Moreover, the combination of 
experimental data on $\sigma$ and $\sigma_{\rm FB}$ at different values of 
the c.m.\ energy allows to further restrict such separate bounds in a 
model-independent way.
\par
To clarify this statement and intuitively show by a simplified example 
the role of the different energy points in improving the constraints, 
assuming that no deviation from the SM is observed within the experimental 
accuracies, constraints on the contact interaction couplings  
$\epsilon_{\alpha\beta}$ can be derived from the system of two inequalities:
\begin{equation}
\vert\sigma^{\rm SM+CI}-\sigma^{\rm SM}\vert < \delta\sigma,
\label{delsig}
\end{equation}
\begin{equation}
\vert A_{\rm FB}^{\rm SM+CI}-A_{\rm FB}^{\rm SM}\vert < \delta A_{\rm FB},
\label{delafb}
\end{equation}
where $\delta\sigma$ and $\delta A_{\rm FB}$ represent the experimental 
uncertainties on these observables. Taking Eqs.~(\ref{s+}) and (\ref{s-}) 
into account, the deviations from the SM predictions in the left-hand sides 
of Eqs.~(\ref{delsig}) and (\ref{delafb}) can be written as: 
\begin{equation}
\sigma^{\rm SM+CI}-\sigma^{\rm SM}=\frac{1}{4}
\left[\left(\Delta\sigma_{\rm LL}+\Delta\sigma_{\rm RR}\right)+
\left(\Delta\sigma_{\rm LR}+\Delta\sigma_{\rm RL}\right)\right],
\label{ddelsig}
\end{equation}
\begin{equation}
A_{\rm FB}^{\rm SM+CI}-A_{\rm FB}^{\rm SM}=
\frac{3}{16\hskip2pt\sigma^{\rm SM}}
\left[\left(1-\frac{4}{3}A_{\rm FB}^{\rm SM}\right)
\left(\Delta\sigma_{\rm LL}+\Delta\sigma_{\rm RR}\right)
-\left(1+\frac{4}{3}A_{\rm FB}^{\rm SM}\right)
\left(\Delta\sigma_{\rm LR}+\Delta\sigma_{\rm RL}\right)\right],
\label{ddelafb}
\end{equation}
where $\Delta\sigma_{\alpha\beta}=
\sigma_{\alpha\beta}^{\rm SM+CI}-\sigma_{\alpha\beta}^{\rm SM}$.


From Eqs.~(\ref{delsig})--(\ref{ddelafb}) one can obtain constraints on 
the $\epsilon_{\alpha\beta}$. Specifically, the areas allowed to the 
values of the parameters are enclosed by concentric circles in the 
planes $(\epsilon_{LL},\epsilon_{RR})$ and $(\epsilon_{LR},\epsilon_{RL})$. 
For example, the domain allowed to the pair 
$(\epsilon_{\rm LL},\epsilon_{\rm RR})$ is delimited by the circular 
contours: 
\begin{equation}
\left(\epsilon_{\rm LL}
+\frac{\alpha_{\rm e.m.}}{s}{\cal M}_{\rm LL}^{\rm SM}\right)^2
+\left(\epsilon_{\rm RR}
+\frac{\alpha_{\rm e.m.}}{s}{\cal M}_{\rm RR}^{\rm SM}\right)^2
=R_{\pm}^2,
\label{circ}
\end{equation}
where 
\begin{equation}     \label{Eq:radius}
R_{\pm}^2
=\left(\frac{\alpha_{\rm e.m.}}{s}{\cal M}_{\rm LL}^{\rm SM}\right)^2 
+\left(\frac{\alpha_{\rm e.m.}}{s}{\cal M}_{\rm RR}^{\rm SM}\right)^2
\pm\kappa^2,
\end{equation}
and 
\begin{equation}
\kappa^2=\left(\frac{\alpha_{\rm e.m.}}{s}\right)^2
\frac{4}{\sigma_{\rm pt}}\delta\sigma_{+}.
\label{circ1}
\end{equation}
In the right-hand side of Eq.~(\ref{circ1}), $\delta\sigma_{+}$ must be
expressed in terms of the experimental uncertainties $\delta\sigma$ and
$\delta A_{\rm FB}$,\footnote{While $\sigma_+$ and $\sigma_-$, 
as shown in Eqs.~(\ref{Eq:sigma+})
and (\ref{Eq:sigma-}), are the most natural observables, we use instead
$\sigma$ and $A_{\rm FB}$ for our analysis,
since the data are tabulated (with errors) for these observables.}
and a color factor $1/N_C$ is needed in the case of
quark-antiquark production.  These relations show that both the centre and the
radii of the circles $R_\pm$ are determined by the values of the SM helicity
amplitudes and depend on energy, while the width of the allowed area is
determined by the experimental uncertainty of the observables.  Therefore, in
principle, the combination of two (or more) such allowed regions corresponding
to different energies can lead to a reduction of the allowed region and,
ultimately, to model-independent bounds on the contact interaction coupling
constants.

It should be stressed that, while the observables are given by {\it two
sums} of helicity cross sections, $\sigma_{\rm LL}+\sigma_{\rm RR}$
and $\sigma_{\rm LR}+\sigma_{\rm RL}$, it does {\it not} follow that one can
only obtain constraints on the corresponding sums of parameters,
$\epsilon_{\rm LL}+\epsilon_{\rm RR}$ and 
$\epsilon_{\rm LR}+\epsilon_{\rm RL}$. There is indeed a small, but finite,
sensitivity to the individual parameters.
This is due to effects proportional to squares of the $\epsilon$ parameters,
together with the small difference between left- and right-handed couplings
of the Standard Model, $|g_{\rm L}^\ell|\ne|g_{\rm R}^\ell|$.

\section{Data fitting and derivation of constraints}
Recently, the $f\bar f$ Subgroup of LEPEWWG presented preliminary combined
results of measurements of the four LEP collaborations using experimental data
from the full LEP2 available data set at energies from 130~GeV up to 207~GeV
for the annihilation processes $e^+e^- \to f \bar f$
\cite{geweniger}. In particular, for lepton final states
$f=\mu$ and $\tau$, the set of the average cross sections $\sigma _{\mu \mu}$,
$\sigma_{\tau \tau}$ and forward-backward asymmetries $A_{\rm FB}^{\mu \mu}$,
$A_{\rm FB}^{\tau \tau}$ and their experimental errors have been given for 
the 12 energy points listed in Table~\ref{table:luminosity}.

\begin{table}[h]
\centering
\caption{
Approximate average integrated luminosity per experiment 
and nominal centre-of mass energies collected during LEP2 operations 
\cite{geweniger}.}
\vspace*{8pt}
\begin{tabular}{|l|c|c|c|c|c|c|c|c|c|c|c|c|}
\hline
\hline
 $E_{CM}$ (GeV) 
&  130 & 136 & 161 & 172 & 183 & 189 & 192 & 196 & 200 & 202 & 205 & 207 \\
$\Lumint$ $[{\rm pb}^{-1}]$ 
&    3 &   3 &  10 &  10 &  50 & 170 &  30 &  80 &  80 &  40 &  80 & 140 \\ 
\hline\hline
\end{tabular} 
\label{table:luminosity}
\end{table}

The data fitting procedure used is based on the method of least squares.
We introduce a $\chi^2$ function, which may be written in the following 
matrix form
\begin{equation}
\chi^2(\boldsymbol{ \epsilon})=({\boldsymbol{\cal  O}}^{\rm LEP2} 
- {\boldsymbol{\cal O}}^{\rm TH}(\boldsymbol{ \epsilon}))^{\rm T}
V^{-1}(\boldsymbol{\cal O}^{\rm LEP2} - \boldsymbol{\cal O}^{\rm TH}
(\boldsymbol{\epsilon})),
\end{equation}
where $\boldsymbol{\epsilon}=(\epsilon_{\rm RR},\epsilon_{\rm LL},
\epsilon_{\rm RL},\epsilon_{\rm LR})$ is the vector of C.I. parameters;
${\boldsymbol{\cal O}}^{\rm LEP2}$ is the vector of values of observables 
measured at LEP2 and $\boldsymbol{\cal O}^{\rm TH}$ is the vector of their 
theoretical predictions; finally, $V$ is the covariance matrix
of the experimental uncertainties.
\par 
The chosen set of observables, represented by the vector
$\boldsymbol{\cal O}^{\rm LEP2}$, contains 48 elements (two kinds of
observable for two flavour channels and twelve energy points). The
corresponding theoretical predictions, $\boldsymbol{\cal O}^{\rm TH}$,
which depend on the CI parameters $\boldsymbol{\epsilon }$
and on radiative corrections via
improved Born SM amplitudes \cite{Consoli:1989pc,Altarelli:1990dt}, 
have been evaluated with $m_{\rm top}=175$~GeV and $m_H=150$~GeV. 
Initial- and final-state radiation are taken into account by the program 
ZFITTER \cite{Bardin:1999yd} adapted to the present case of contact 
interactions.  
The radiative corrections were applied using definition ``2'' in
Ref.~\cite{geweniger}: namely, for dilepton events $\sqrt{s'}$ is taken
to be the bare invariant mass of the outgoing dilepton pair (as opposed to
that of the $s$-channel propagator), the ISR-FSR photon interference is
included and the signal is defined by the kinematical cut $\sqrt{s'} >
0.85\sqrt{s}$.
We note that the improved Born amplitudes leave the form of the previous 
equations for the cross sections, Eqs.~(\ref{cross}) and (\ref{sig+-}) 
unaltered.
\par 
As regards the $48 \times 48$ symmetric covariance matrix $V$, the diagonal 
entries are the experimental uncertainties on the observables, while 
the off-diagonal entries define the correlations between the observables 
as well as among the different energy points \cite{geweniger}.
\par
The least-square confidence region is determined by the condition
\begin{equation}
\chi^2(\boldsymbol{\epsilon}) \leq \chi^2_{\rm min} + \chi^2_{\rm CL},
\label{chi2}
\end{equation}
where $\chi^2_{\rm min}$ is the minimum value of the function
$\chi^2(\boldsymbol{\epsilon})$ and $\chi^2_{\rm CL} = 9.49$ for $95~\%$ CL
and four degrees of freedom. The procedure of minimization
$\chi^2(\boldsymbol{\epsilon})$ is performed using the program package MINUIT
\cite{James:1975dr}. 
\begin{table}[htb]
\centering
\caption{Central value $\epsilon^0$, global limits (allowed
intervals) obtained as projections of the 95\% CL four-dimensional region on
the axes and 95\% CL one-dimensional model-dependent constraints on the CI
parameters.}
\vspace*{8pt}
\setlength{\extrarowheight}{6pt}
\begin{tabular}{|c|c|c|c|}
\hline
 Parameter & \multicolumn{2}{c|}{Model independent}
            & Model dependent\\ \cline{2-3}
 $[\text{TeV}^{-2}]$ & central value & global limits & \\ \hline \hline
$\epsilon_{\rm LL}$ &  $\phantom{-}0.0085$ & ($-0.175$, 0.095) 
& $-0.0047^{+0.0071}_{-0.0071}$ \\[4pt]
\hline
$\epsilon_{\rm RR}$ & $-0.0195$ & ($-0.187$, 0.111) 
& $-0.0052^{+0.0078}_{-0.0078}$ \\[4pt]
\hline
$\epsilon_{\rm LR}$ &  $\phantom{-}0.0120$ & ($-0.225$, 0.060) 
& $-0.0012^{+0.0111}_{-0.0116}$ \\[4pt]
\hline
$\epsilon_{\rm RL}$ & $-0.0160$ & ($-0.225$, 0.060) 
& $-0.0012^{+0.0111}_{-0.0116}$ \\[4pt]
\hline
\end{tabular}
\label{table:epsilon}
\end{table}

Combining the $\mu$ and $\tau$ data, we show 
in Table~\ref{table:epsilon} the components of the central value
$\epsilon^0$ (over-all minimum of $\chi^2$) and the global limits (intervals
$(\epsilon_{\rm min},\epsilon_{\rm max})$) obtained as projections of the
confidence region on the corresponding axes. These intervals should be
considered as global, model-independent, constraints on the CI parameters
$\epsilon_{\alpha \beta}$.  The $\chi^2$ in the model-independent fits
amounted to 41.3, for $n_d = 48 - 4 = 44$ degrees of freedom: the probability
of this result is $p=0.411$\cite{Hagiwara:pw}.  For comparison, we give the
95\% CL one-parameter constraints on $\epsilon_{\alpha \beta}$ parameters for
the LL, RR, LR and RL contact-interaction models.

In Figs.~\ref{fig1}--\ref{fig2} we show the contours which
bound the regions found as ``projections'' of the four-dimensional confidence
hypervolume determined by (\ref{chi2}) on four of the two-dimensional 
planes (LL--RR), (LR--RL), (LL--LR), (LL--RL). 
The contours have been produced as the line connecting all points of the plane 
where $\chi^2$ takes the value $\chi^2_{\rm min} + \chi^2_{\rm CL}$ after 
minimization on the two remaining free parameters.

\begin{figure}[htb]
\refstepcounter{figure}
\label{fig1}
\addtocounter{figure}{-1}
\begin{center}
\setlength{\unitlength}{1cm}
\begin{picture}(12,8)
\put(-1.5,0.0)
{\mbox{\epsfysize=8cm\epsffile{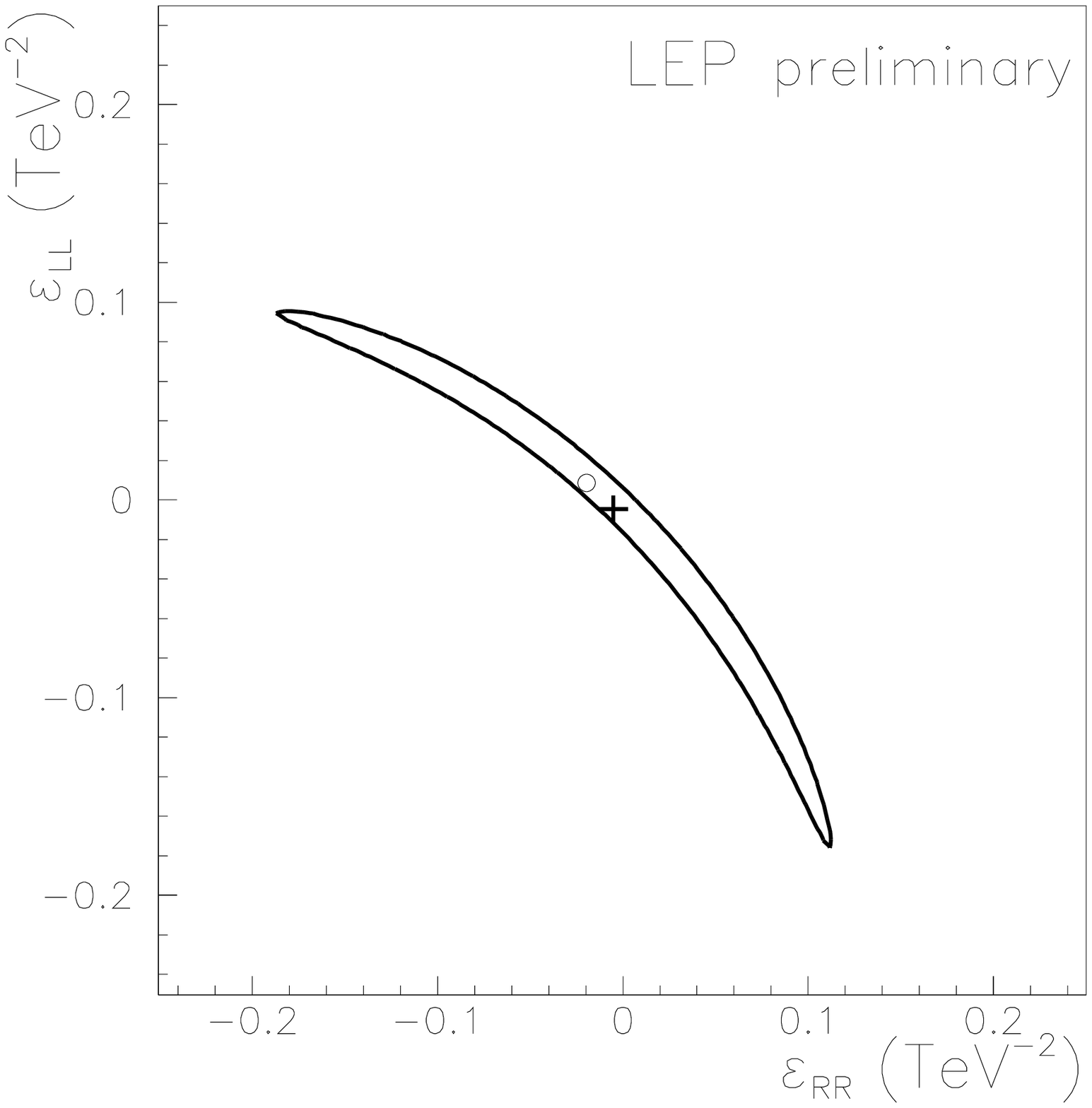}}
 \mbox{\epsfysize=8cm\epsffile{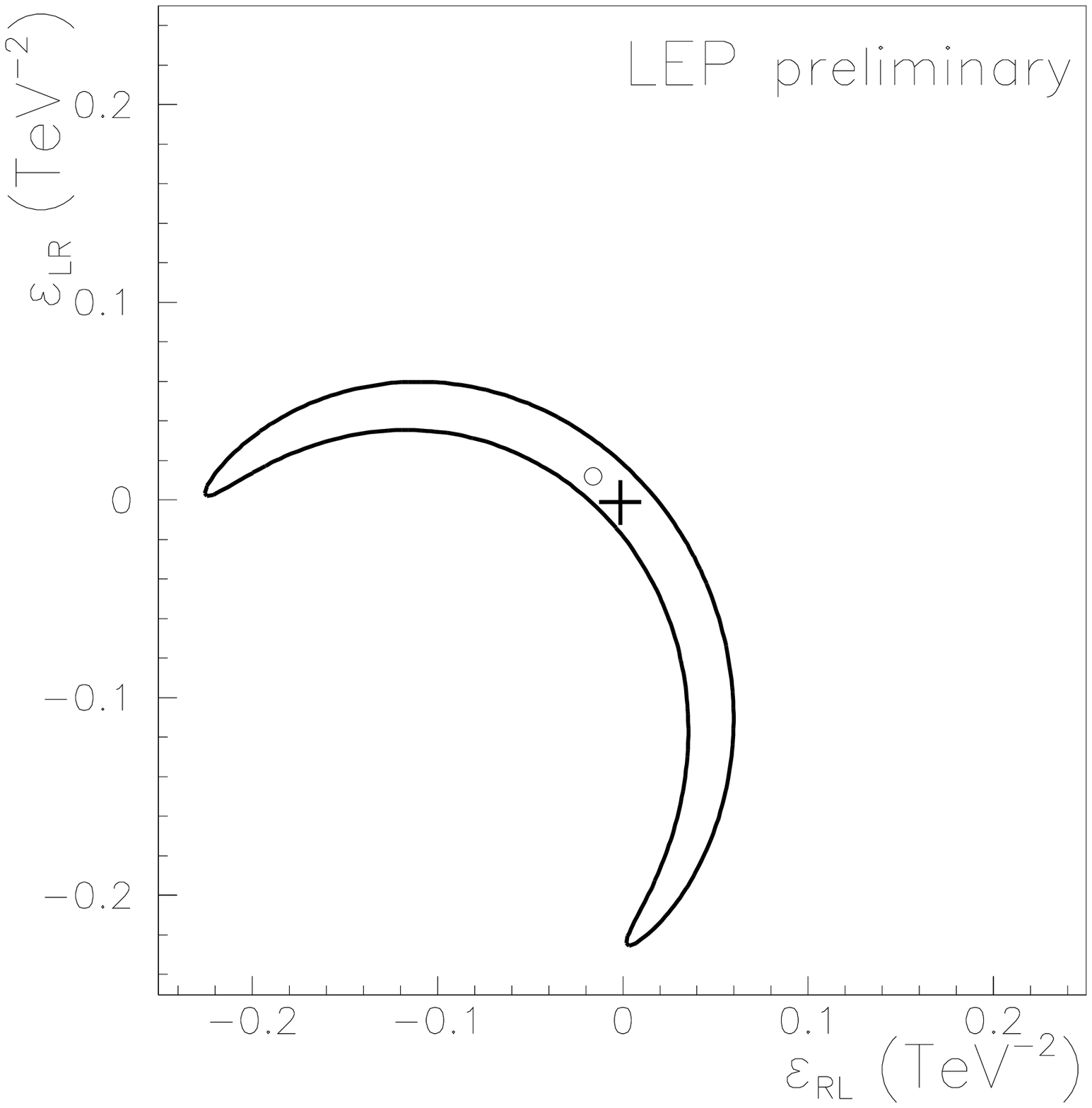}}}
\end{picture}
\vspace*{-3mm}
\caption{Allowed areas at 95\% C.L. on leptonic contact interaction 
parameters in the planes ($\epsilon_{\rm LL},\epsilon_{\rm RR}$) and
($\epsilon_{\rm LR},\epsilon_{\rm RL})$, obtained as projections of
the four-dimensional confidence hypervolume on the relevant plane after
minimization in the two remaining parameters. The bars correspond to 
one-dimensional model-dependent constraints as discussed in the text.
The circles correspond to the central values (see Table~\ref{table:epsilon}).}
\end{center}
\end{figure}
\begin{figure}[htb]
\refstepcounter{figure}
\label{fig2}
\addtocounter{figure}{-1}
\begin{center}
\setlength{\unitlength}{1cm}
\begin{picture}(12,8)
\put(-1.5,0.0)
{\mbox{\epsfysize=8cm\epsffile{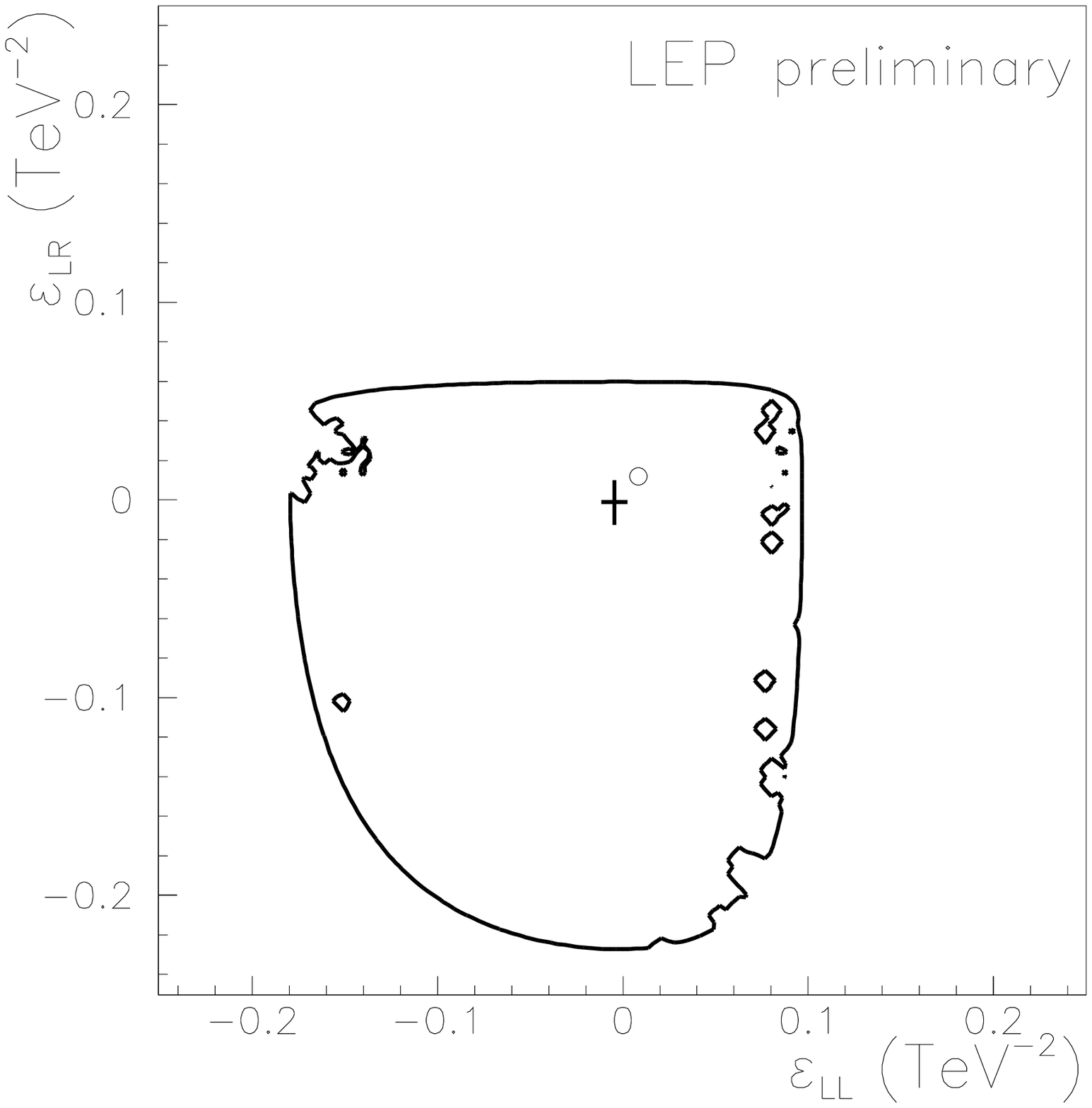}}
 \mbox{\epsfysize=8cm\epsffile{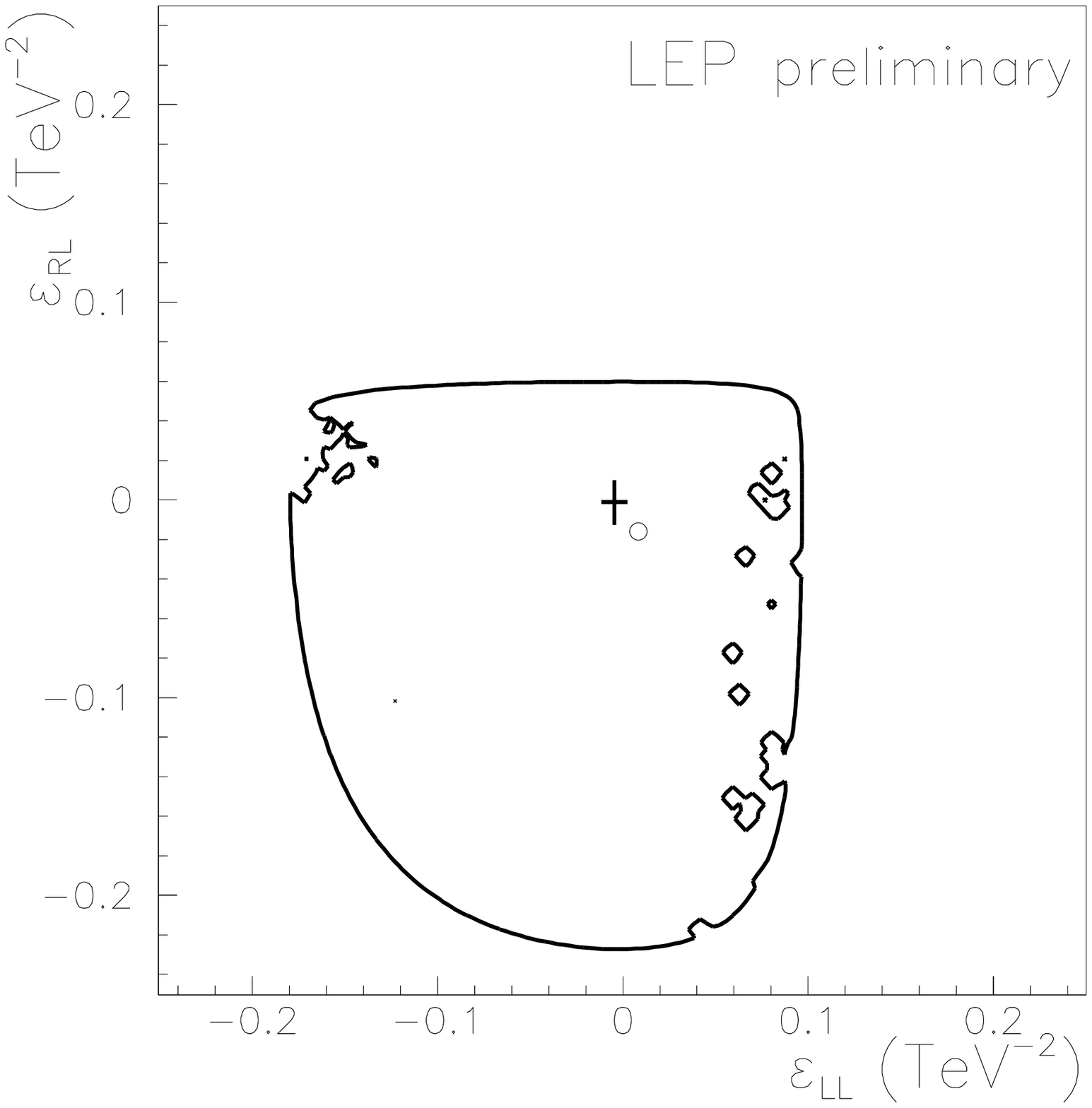}}}
\end{picture}
\vspace*{-3mm}
\caption{The same as Fig.~1 for the ($\epsilon_{\rm LL},\epsilon_{\rm LR}$) 
and ($\epsilon_{\rm LL},\epsilon_{\rm RL}$) planes.}
\end{center}
\end{figure}

These figures show obvious symmetries.
First of all, in Fig.~\ref{fig1}, where we display the allowed regions
in the ($\epsilon_{\rm LL},\epsilon_{\rm RR}$) and
($\epsilon_{\rm LR},\epsilon_{\rm RL})$ planes,
there is an approximate ``reflection symmetry'' between 
$\epsilon_{\rm LL}\leftrightarrow\epsilon_{\rm RR}$
as well as between 
$\epsilon_{\rm LR}\leftrightarrow\epsilon_{\rm RL}$.
As discussed in Sec.~2, the observables depend on
$\sigma_{\rm LL}+\sigma_{\rm RR}$ and
$\sigma_{\rm LR}+\sigma_{\rm RL}$, and to lowest order
in the $\epsilon$'s, this translates into a dependence
on $\epsilon_{\rm LL}+\epsilon_{\rm RR}$ and
$\epsilon_{\rm LR}+\epsilon_{\rm RL}$.
Thus, in this approximation, the allowed regions would be bands
at fixed $\epsilon_{\rm LL}+\epsilon_{\rm RR}$ and
$\epsilon_{\rm LR}+\epsilon_{\rm RL}$, representing strong
correlations between pairs of parameters.
The contributions of the second-order effects 
(in the $\epsilon_{\alpha\beta}$) delimit and curve these bands.
In the case of the $\epsilon_{\rm LR}$--$\epsilon_{\rm RL}$
exclusion region, the radius of curvature,
given by an expression analogous to Eq.~(\ref{Eq:radius}), is smaller
than that of the $\epsilon_{\rm LL}$--$\epsilon_{\rm RR}$ exclusion region.
This stronger bending originates from the destructive {\it vs.}
constructive interference between photon- and $Z$-exchange:
above the $Z$ resonance, $g_{\rm L}^\ell g_{\rm R}^\ell\chi_Z<0$,
whereas $(g_{\rm L}^\ell)^2\chi_Z\approx(g_{\rm R}^\ell)^2\chi_Z>0$, so
$|{\cal M}_{\rm LR}^{\rm SM}|=|{\cal M}_{\rm RL}^{\rm SM}|
< |{\cal M}_{\rm LL}^{\rm SM}|\approx|{\cal M}_{\rm RR}^{\rm SM}|$.

In Fig.~\ref{fig2}, we show the analogous allowed regions in the 
($\epsilon_{\rm LL},\epsilon_{\rm LR}$) 
and ($\epsilon_{\rm LL},\epsilon_{\rm RL}$) planes.
The allowed regions in the ($\epsilon_{\rm RR},\epsilon_{\rm LR}$) 
and ($\epsilon_{\rm RR},\epsilon_{\rm RL}$) planes are very similar 
to those of this Fig.~\ref{fig2}, and hence not shown.
\par 
In the figures, the constraints on the one-parameter models LL, RR 
LR and RL (see Table~\ref{table:epsilon}) are represented by bars.
These correspond to one-dimensional model-dependent constraints 
at 95 \% C.L. with $\chi^2_{\rm CL} = 3.84 $, obtained by varying 
only one parameter at a time with the remaining three set equal zero.
The results in this case are in full agreement with those obtained in 
Ref.~\cite{geweniger}.
\par 
Figure \ref{fig2} is rather different
from Fig.~\ref{fig1}, but the two panels are very similar among themselves.
This is due to the symmetric inputs,
$\sigma_{\rm LL}+\sigma_{\rm RR}$ and $\sigma_{\rm LR}+\sigma_{\rm RL}$,
together with the fact that the linear approximation, determined by the 
interference between SM and CI couplings, provides a first,
rough description of the bounds.
Also, we note that there is little correlation between these pairs
of parameters, and the allowed regions are simply
\begin{equation}
[\epsilon_{\alpha\beta}<\delta']\cap
[\epsilon_{\alpha'\beta'}<\delta']\cap
[\epsilon_{\alpha\beta}^2+\epsilon_{\alpha'\beta'}^2<\delta^2], 
\end{equation}
where
\begin{equation}
\delta'\simeq\text{max}(\epsilon_{\rm LL})\quad\text{or}\quad
             \text{max}(\epsilon_{\rm LR})\quad\text{or}\quad
             \text{max}(\epsilon_{\rm RL}),
\end{equation}
as determined from Fig.~\ref{fig1}.
Similarly,
\begin{equation}
\delta^2=[\text{max}(\epsilon_{\rm LL})]^2
+\{[\text{max}(\epsilon_{\rm LR})]^2\quad\text{or}\quad
   [\text{max}(\epsilon_{\rm RL})]^2\},
\end{equation}
respectively, for the two panels. This simple shape is thus due to
the lack of correlations among the parameters shown in Fig.~\ref{fig2}.

\section{Discussion}
Our most important result is that if one does not restrict the analysis to
individual models, the bounds on the $\epsilon$'s are rather loose.
In fact, any set of three of them (but not all four, as is seen from
the correlations in Fig.~\ref{fig1}) can be of the order of 
$0.2~{\rm TeV}^{-2}$. This corresponds to a scale $\Lambda\sim2.2~{\rm TeV}$.
\par
In the case of $\epsilon_{\rm LL}$ and $\epsilon_{\rm RR}$, as discussed
above, the orientation of the ``banana'' in Fig.~\ref{fig1} implies that
$\epsilon_{\rm LL}$ and $\epsilon_{\rm RR}$ should roughly add to zero. Rather
large deviations from the SM are allowed, {\it provided these parameters have
opposite signs}. It should be noted that, if one assumes universality, such
opposite signs can never arise from the low-energy limit of a vector-particle
exchange, irrespective of the chiralities of the couplings.  This may in part
explain why the present bounds are much looser than those of the
model-dependent analyses.

Also, it should be stressed that we do not assume full lepton universality
in this analysis. The muon and tau data are combined,
but the couplings to those currents are not taken to be the same as
the couplings to the electron of the initial state.
If lepton universality were imposed, one would have the additional constraint
$\epsilon_{\rm LR}=\epsilon_{\rm RL}$.
Also, full lepton universality would imply the product
$\epsilon_{\rm LL}\,\epsilon_{\rm RR}>0$, and much of the allowed
part of Fig.~\ref{fig1} would be excluded.
We note that there are models without flavour universality,
where $\epsilon_{\rm LL}$ and $\epsilon_{\rm RR}$ can have opposite signs
(see, e.g.\ \cite{Gounaris:1997ft,Lynch:2000md}).

\begin{figure}[htb]
\refstepcounter{figure}
\label{fig-tau}
\addtocounter{figure}{-1}
\begin{center}
\setlength{\unitlength}{1cm}
\begin{picture}(12,8)
\put(-1.5,0.0)
{\mbox{\epsfysize=8cm\epsffile{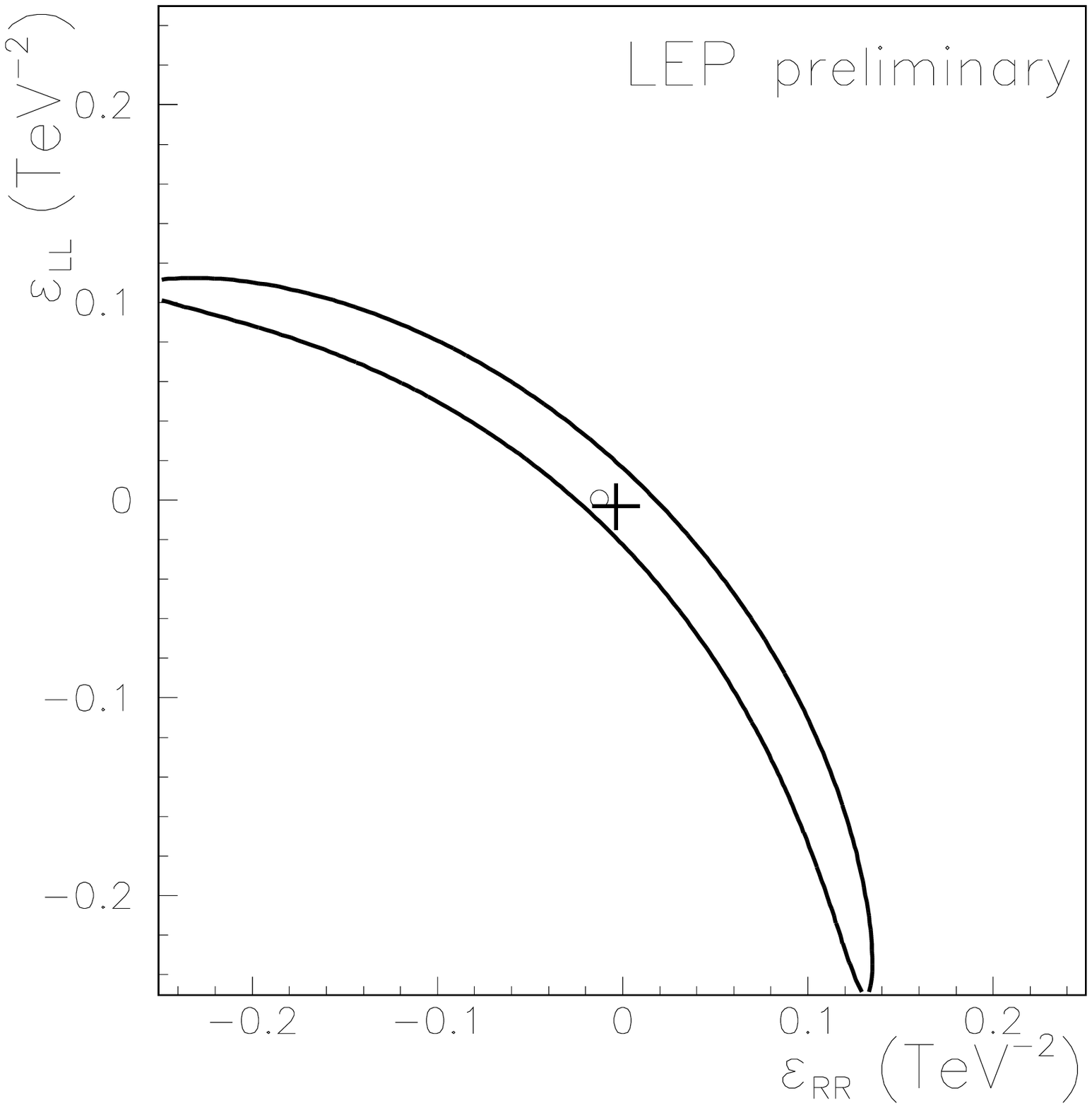}}
 \mbox{\epsfysize=8cm\epsffile{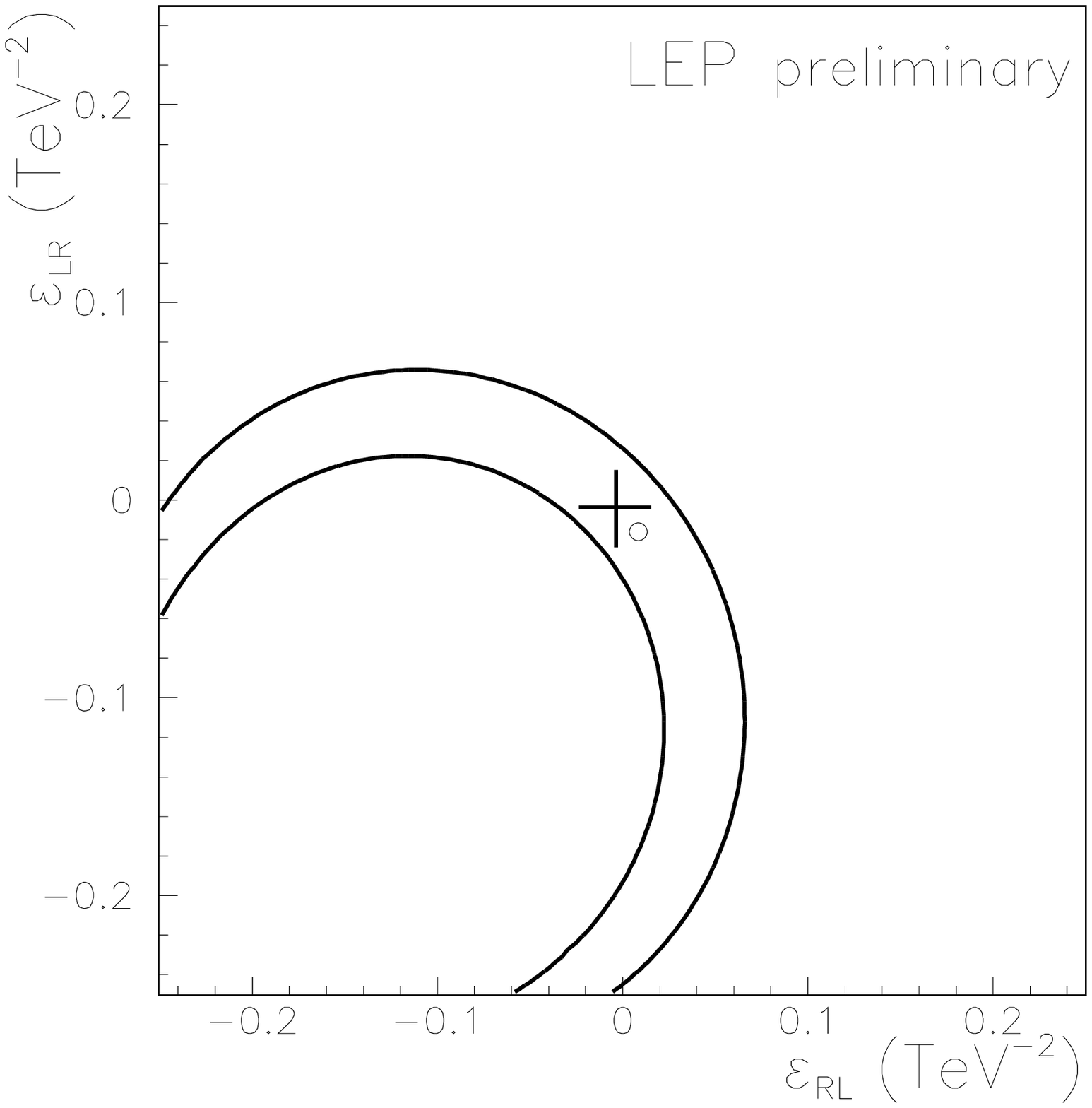}}}
\end{picture}
\vspace*{-3mm}
\caption{Similar to Fig.~\ref{fig1}, for tau data only.}
\end{center}
\end{figure}

Up to this point, we have combined the muon and tau data.  It is also
interesting to study these two data sets separately.  In Fig.~\ref{fig-tau} we
show the contours which bound the allowed regions in the $(\epsilon_{\rm
LL},\epsilon_{\rm RR})$ and $(\epsilon_{\rm LR},\epsilon_{\rm RL})$ planes for
$e^+e^-\to \tau^+\tau^-$ (i.e., without using the muon data).  The general
shapes of these allowed regions are rather similar to those obtained from the
combined data, but they are significantly larger.  The corresponding allowed
intervals of the $\epsilon$ parameters are given in
Table~\ref{table:epsilon-tau}, the analogue of Table~\ref{table:epsilon}.  The
model-independent global limits, for example, are looser than the combined
muon and tau analysis by up to $40\%$. The muon data alone give shapes and
allowed intervals quite similar to those in Fig.~\ref{fig-tau} and
Table~\ref{table:epsilon-tau}, respectively, but narrower, essentially
reflecting the larger total error in the $\tau$ case.
\begin{table}[htb]
\centering
\caption{Similar to Table~\ref{table:epsilon}, for final-state $\tau$ pairs
only.}
\vspace*{8pt}
\setlength{\extrarowheight}{6pt}
\begin{tabular}{|c|c|c|c|}
\hline
 Parameter & \multicolumn{2}{c|}{Model independent}
            & Model dependent\\ \cline{2-3}
 $[\text{TeV}^{-2}]$ & central value & global limits & \\ \hline \hline
$\epsilon_{\rm LL}$ &  $\phantom{-}0.0005$ & ($-0.249$, 0.113) 
& $-0.0032^{+0.0118}_{-0.0120}$ \\[4pt]
\hline
$\epsilon_{\rm RR}$ & $-0.0125$ & ($-0.258$, 0.136) 
& $-0.0035^{+0.0129}_{-0.0131}$ \\[4pt]
\hline
$\epsilon_{\rm LR}$ &  $-0.016$ & ($-0.273$, 0.066) 
& $-0.0036^{+0.0188}_{-0.0202}$ \\[4pt]
\hline
$\epsilon_{\rm RL}$ & $\phantom{-}0.0085$ & ($-0.273$, 0.066) 
& $-0.0036^{+0.0188}_{-0.0202}$ \\[4pt]
\hline
\end{tabular}
\label{table:epsilon-tau}
\end{table}

In specific models, there are often constraints on these deviations
$\epsilon_{\alpha\beta}$. For example, the $Z'$ couplings of $E_6$ models
lead to the constraints \cite{Leike:1998wr}:
\begin{align}
Z'_\chi:\quad&\epsilon_{\rm LR}=\epsilon_{\rm RL}<0, \quad
\epsilon_{\rm LL}=9\,\epsilon_{\rm RR}<0, \\
Z'_\psi:\quad&\epsilon_{\rm LR}=\epsilon_{\rm RL}>0, \quad
\epsilon_{\rm LL}=\epsilon_{\rm RR}<0, \\
Z'_\eta:\quad&\epsilon_{\rm LR}=\epsilon_{\rm RL}<0, \quad
\epsilon_{\rm LL}=\frac{1}{4}\,\epsilon_{\rm RR}<0.
\end{align}
These signs are given by the signs of the couplings, together with
the low-energy limit of the propagator.
The leptonic data studied here, lead to the $M_{Z'}$ bounds:
$Z'_\chi$: $600~\text{GeV}$; $Z'_\psi$: $330~\text{GeV}$; $Z'_\eta$: 
$340~\text{GeV}$.
The corresponding bounds from all data \cite{geweniger} are
670~GeV, 480~GeV and 430~GeV, respectively.

Also, in the case of models with TeV-scale extra dimensions (with
Kaluza--Klein excitations of the photon and the $Z$), there are relations
among the couplings \cite{Pasztor:2001hc}: $\epsilon_{\rm LR}=\epsilon_{\rm
RL}<0$, and $\epsilon_{\rm LL}=\epsilon_{\rm RR}/4s_W^2\,\simeq\epsilon_{\rm
RR}<0$.  For one extra dimension, the bound on the compactification scale 
\cite{Rizzo:1999br} is $M_c>2.2~\text{TeV}$.
\begin{figure}[htb]
\refstepcounter{figure}
\label{fig3}
\addtocounter{figure}{-1}
\begin{center}
\setlength{\unitlength}{1cm}
\begin{picture}(12,8)
\put(1,0.0)
{\mbox{\epsfysize=8cm\epsffile{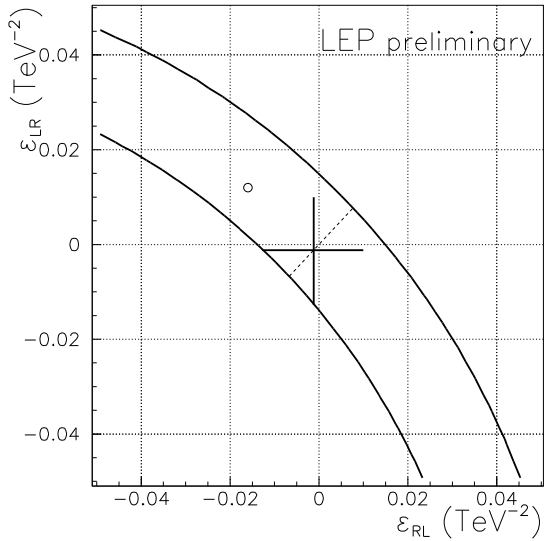}}}
\end{picture}
\vspace*{-3mm}
\caption{Similar to Fig.~\ref{fig1} (right panel), but
magnification of the region of small $\epsilon_{\rm LR}$
and $\epsilon_{\rm RL}$, for $\chi^2_{\rm CL}=5.99$.
The diagonal line corresponds to the constraint of the anomalous gauge 
coupling model.}
\end{center}
\end{figure}

As anticipated in the Introduction, we shall here consider an example
application, namely the effects of anomalous gauge couplings
\cite{Gounaris:1997ft} in the process (\ref{proc}). We note that this model,
which assumes universality, is characterized by two parameters, $f_{DB}$ and
$f_{DW}$.  The deviations (\ref{amplit}) will take the form
\begin{equation}  \label{Eq:agc}
\epsilon_{\rm LL}=\alpha_{e.m.}
\left(\frac{\tilde f_{DW}}{2s_W^2}+\frac{2 \tilde f_{DB}}{c_W^2}\right),\quad
\epsilon_{\rm RR}=\alpha_{e.m.}\frac{8\tilde f_{DB}}{c_W^2}, \quad
\epsilon_{\rm LR}=\epsilon_{\rm RL}
=\alpha_{e.m.}\frac{4 \tilde f_{DB}}{c_W^2}, 
\end{equation}
where $\tilde f_{DB}$ and $\tilde f_{DW}$ are related to $f_{DB}$ and $f_{DW}$
of ref.~\cite{Gounaris:1997ft} by $\tilde f=f/m_t^2$.  In this model, one has
$\epsilon_{\rm LR}=\epsilon_{\rm RL}$, so any deviation would be restricted to
lie along the dashed line in Fig.~\ref{fig3}, which shows a magnification of
the allowed band in Fig.~\ref{fig1} (right panel) for $\chi_{\rm CL}^2=5.99$,
corresponding to two parameters. The intersections with the allowed bounds
allow us to set a limit on $|\tilde f_{DB}|< 0.21~\text{TeV}^{-2}$.  This also
amounts to a bound on $\epsilon_{\rm RR}$. From the analogue of
Fig.~\ref{fig1} (left panel), corresponding to two parameters, one can extract
bounds on $\epsilon_{\rm LL}$. Using Eq.~(\ref{Eq:agc}), these can then be
converted to the bounds: $-1.7~\text{TeV}^{-2}<\tilde
f_{DW}<1.1~\text{TeV}^{-2}$.

We have not analyzed the quark data, which is of poorer quality, due
to the limited efficiency of $b$-tagging, together with the problem
of distinguishing $b$ from $\bar b$ jets (see, however, 
Ref.~\cite{Inoue:2000hc}).

At the Linear Collider, where polarization would be available, more
observables can be studied, such as $A_{\rm LR}$ and $A_{\rm LR, FB}$. Thus,
and because of the higher energy, dramatically tighter constraints are
foreseen \cite{Babich:2001nc}.

\bigskip
\medskip
\leftline{\bf Acknowledgements} 
\par\noindent 
This research has been supported by the Research Council of Norway, by MIUR
(Italian Ministry of University and Research) and by funds of the University
of Trieste.  We would like to thank our colleagues of the Alice, BaBar and
Compass collaborations for the use of their computing facilities, which
significantly sped up our work.

\end{document}